\documentclass[letterpaper,twocolumn, aps, 10pt, pra, floatfix]{revtex4-1}
\usepackage{epsfig}
\usepackage{amsmath}
\usepackage{amssymb}
\usepackage{graphicx}
\usepackage{indentfirst}
\usepackage[tight]{subfigure}
\usepackage[latin1]{inputenc}
\usepackage{float}

\DeclareMathOperator{\sech}{sech}

\begin{document}

\title{Dissipation-driven squeezed and sub-Poissonian mechanical states in quadratic optomechanical systems }
\author{T. Figueiredo Roque}
\email{tfroque@ifi.unicamp.br}
\author{A. Vidiella-Barranco}
\email{vidiella@ifi.unicamp.br}
\affiliation{Instituto de Física Gleb Wataghin, Universidade Estadual de Campinas, 13083-859 Campinas, São Paulo, Brazil}
\date{\today}

\begin{abstract}
In this work we study an optomechanical system in which there is a purely quadratic optomechanical coupling between the optical and mechanical modes. The optical mode is pumped by three coherent fields and the mechanical mode is parametrically driven. We show that if the frequencies and amplitudes of both optical and mechanical drivings are properly chosen, the optomechanical interaction gives rise to an effective interaction, which, in the presence of optical damping and in the absence of mechanical damping, has the squeezed vacuum state and the squeezed one phonon state as dark states of the dynamics. These states are well known for presenting quadrature squeezing and sub-Poissonian statistics. However, even in the presence of mechanical damping it is possible to find steady states with large degrees of quadrature squeezing or strong sub-Poissonian statistics. 
\end{abstract}
\maketitle

\section*{I. INTRODUCTION}
The observation of quantum features in macroscopic mechanical systems is still a big challenge. The reason behind this lies essentially on the fact that the interaction of macroscopic systems with the environment induces strong decoherence processes, so that quantum phenomena are usually masked. However, recent experimental progress in quantum optomechanics, the area of physics that deals with systems in which optical and mechanical degrees of freedom can interact, brings good perspectives. The experimental realization of sideband cooling of macroscopic mechanical oscillators \cite{arcizet, gigan, schliesser, thompson, teufel, rocheleau,teufel3} and strong coupling between an optical field and a mechanical oscillator \cite{groblacher, teufel2} are specially promising. As a result of such remarkable experimental achievements, a number of theoretical studies proposing the preparation of nonclassical states in such systems appeared, including entangled states, quantum superposition states, sub-Poissonian states and squeezed states.

The generation of squeezed states is specially important not only because of the theoretical interest in such states, but also because of their technological applications, like in the development of ultrasensitive force sensors \cite{caves} and applications in quantum information processing using continuous variable states \cite{braunstein}. Among the several schemes proposed to prepare those states, we observe two major approaches: the first one relies on the fact that the optomechanical coupling induces an effective spring constant in the mechanical oscillator, the so called optical spring effect. Therefore, it is possible to induce a time-dependent effective spring constant, generating the desired squeezing of the mechanical oscillator \cite{mari, asjad, schmidt, liao}. In such proposals, however, usually small degrees of squeezing are reached (slightly above $3$ $dB$). The second approach uses the fact that right after a measurement the uncertainty in the measured observable is zero, and combines a sequence of quantum measurements and feedback to force the mechanical oscillator into a squeezed state \cite{clerk, ruskov, szorkovsky, szorkovsky2, caves, vanner}. That approach would make possible to reach higher degrees of squeezing, but it is hard to implement with current technology. Other approaches try to squeeze the mechanical oscillator by driving the optical field with squeezed light \cite{jahne}, by parametrically driving a mechanical resonator coupled to a microwave cavity \cite{woolley} or by exploring the quadratic optomechanical coupling \cite{nunnenkamp}. One further way to prepare squeezed states relies on the fact that the optical system usually has a much faster dynamics if compared with the mechanical oscillator, thus, the optical mode can act as an engineered reservoir for the mechanical oscillator \cite{kronwald}. Such scheme is simple and is able to reach high degrees of squeezing, the only experimental limitation being the necessity of working in the resolved sideband regime.

In this work we propose a scheme to prepare the mechanical oscillator in a squeezed vacuum state or in a squeezed one phonon state. Our proposal makes use of an optomechanical system in which there is a purely quadratic coupling between the optical and mechanical modes. The optical mode is pumped by three coherent fields and the mechanical mode is parametrically driven. If the amplitudes and frequencies of both optical and mechanical drivings are properly chosen, the optical field then acts as an engineered reservoir for the mechanical oscillator, being possible to drive it, in the absence of mechanical damping, either to a squeezed vacuum state or to a squeezed one phonon state. Those states are well known for presenting quantum features like quadrature squeezing and sub-Poissonian statistics. Notoriously, the presence of the mechanical damping, although the squeezed vacuum state or the squeezed one phonon state are no longer dark states of the dynamics, it is still possible to reach steady states with high degrees of squeezing or strong sub-Poissonian Statistics. 

The paper is divided as follows: in section II we make a small review about squeezed number states and their properties. In section III we present our model and in section IV we demonstrate how the open dynamics of the system allows us to prepare to target states of the mechanical oscillator. In section V we show our results and section VI is for concluding remarks.

\section*{II. SQUEEZED NUMBER STATES}
The squeezed number states $|\xi, n \rangle$ are defined by the following equation,
\begin{equation}
	|\xi, n \rangle = \hat{S}(\xi) |n \rangle,
\end{equation}
where 
\begin{equation}
	\hat{S}(\xi) = \exp\left[\frac{1}{2} (\xi^{\ast} \hat{a}^2 - \xi \hat{a}^{\dagger 2}) \right]
\end{equation} 
is the squeezing operator, $\xi = |\xi| e^{i \theta}$ is the squeezing parameter, $\hat{a}$ is an annihilation operator and $|n\rangle$ is a number state. The squeezing operator $\hat{S}(\xi)$ is well known in quantum optics for generating squeezed states. Transforming the operator $\hat{a}$ via $\hat{S}(\xi)$, we obtain,
\begin{equation}
	\hat{S}(\xi) \hat{a} \hat{S}^{\dagger}(\xi) = \hat{a} \cosh{|\xi|} + \hat{a}^{\dagger} e^{i \theta} \sinh{|\xi|} = \mu \hat{a} + \nu \hat{a}^{\dagger} = \hat{\beta}(\xi),
	\label{bogoliubov}
\end{equation}
which is the so called Bogoliubov operator, $\hat{\beta}$. For the squeezed vacuum state $|\xi, 0 \rangle$ it can be shown \cite{gerry}, using eq. (\ref{bogoliubov}), that,
\begin{align}
	\langle (\Delta \hat{X}_1)^2 \rangle &= \frac{1}{4} e^{-2 |\xi|}, \label{ux1}\\
	\langle (\Delta \hat{X}_2)^2 \rangle &= \frac{1}{4} e^{2 |\xi|}, \label{ux2}
\end{align} 
where $\hat{X}_1 = (\hat{a} e^{-i \theta/2} + \hat{a}^{\dagger}e^{i \theta/2})/2$ and $\hat{X}_2 = (\hat{a} e^{-i \theta/2} - \hat{a}^{\dagger}e^{i \theta/2})/2 i$ are generalized quadrature operators satisfying the commutation relation $[\hat{X}_1, \hat{X}_2] = i/2$. It is clear from eqs. (\ref{ux1}) and (\ref{ux2}) that the squeezed vacuum state presents quadrature squeezing and that the product $\langle (\Delta \hat{X}_1)^2 \rangle \langle \Delta (\hat{X}_2)^2 \rangle = 1/16$ satisfies the minimum value allowed by the uncertainty principle. For arbitrary squeezed number states $|\xi, n \rangle$ it is possible to show that \cite{kim}, 
\begin{align}
	\langle (\Delta \hat{X}_1)^2 \rangle &= \frac{2 n + 1}{4} e^{-2 |\xi|}, \label{ux1sn}\\
	\langle (\Delta \hat{X}_2)^2 \rangle &= \frac{2 n + 1}{4} e^{2 |\xi|}. \label{ux2sn}
\end{align}
Thus, the states $| \xi, n \rangle$ present quadrature squeezing for $|\xi| > \ln \sqrt{2 n + 1}$. It was shown that squeezed number states also present higher order squeezing \cite{marian}.

To obtain the expansion of $|\xi, 0 \rangle$ in the number state basis, we can proceed this way:
\begin{align}
	&\hat{a} | 0 \rangle  = 0, \\
	&\hat{S}(\xi) \hat{a} \hat{S}^{\dagger}(\xi) \hat{S}(\xi) |0\rangle = 0, \\
	&(\mu \hat{a} + \nu \hat{a}^{\dagger}) |\xi, 0 \rangle = 0.	\label{bv}
\end{align}
Thus, the squeezed vacuum state is the vacuum state of the Bogoliubov annihilation operator $\hat{\beta}(\xi) = \mu \hat{a} + \nu \hat{a}^{\dagger}$. Using the identity $|\xi, 0 \rangle = \sum c_n |n \rangle$ we obtain the following relation between the $c_n$'s,
\begin{equation}
	c_{n+1} = -\frac{\nu}{\mu}\sqrt{\frac{n}{n+1}} c_{n-1},
\end{equation}
for $n \ge 1$, and $c_1 = 0$. Hence, eq. (\ref{bv}) admits one solution, which involves only even number states.

For the squeezed one photon state, its expansion in terms of number states can be obtained in a similar way,
\begin{align}
	&\hat{a}^2 | 1 \rangle  = 0, \\
	&\hat{S}(\xi) \hat{a}^2 \hat{S}^{\dagger}(\xi) \hat{S}(\xi) |1\rangle = 0, \\
	&(\mu \hat{a} + \nu \hat{a}^{\dagger})^2 |\xi, 1 \rangle = 0,\\
	&(\mu^2 \hat{a}^2 + \nu^2 \hat{a}^{\dagger 2} + 2 \mu \nu \hat{a}^{\dagger} \hat{a} + \mu \nu) |\xi, 1 \rangle = 0. \label{bsv}
\end{align}
This means that the squeezed one photon state is the vacuum state of the square of the Bogoliubov annihilation operator, $\hat{\beta}^2(\xi)$. 
Using the identity $|\xi, 1 \rangle = \sum c_n |n \rangle$ we get the following relation between the $c_n$'s,
\begin{align}
	&\sqrt{2} \mu c_2 + \nu c_0 = 0,\\
	&\sqrt{2} \mu c_3 + \sqrt{3} \nu c_1 = 0,\\
	&\sqrt{(n+2) (n+1)} \mu^2 c_{n+2} + \mu \nu (2 n +1) c_n \nonumber \\
	&\qquad \qquad \qquad \quad + \nu^2 \sqrt{n (n-1)} c_{n-2} = 0.
\end{align}
Therefore, eq. (\ref{bsv}) has two linearly independent solutions, one involving only even number states and another one involving only odd number states. Given that a solution of eq. (\ref{bv}) must be also a solution of eq. (\ref{bsv}), the even solution corresponds to the squeezed vacuum state $|\xi, 0 \rangle$. Consequently, the odd solution corresponds to the squeezed one photon state $|\xi, 1 \rangle$.

As pointed out above, squeezed number states $|\xi, n \rangle$ present squeezing if $|\xi| > \ln \sqrt{2 n + 1}$. Nonetheless, these states can also show sub-Poissonian statistics. It can be shown that, for $\theta = 0$,  the second order correlation function of $|\xi, n \rangle$ is given by \cite{kim},
\begin{align}
	g^{(2)}(0) =& 1 - \frac{\cosh(2|\xi|)}{\langle \hat{n} \rangle} n + \frac{\sinh^2{|\xi|}}{\langle \hat{n} \rangle^2} \Bigl[2 n^2 \cosh^2{|\xi|} +  \nonumber\\
	&\qquad \qquad \qquad 2 n \cosh^2{|\xi|} + \cosh(2 |\xi|)\Bigr], \label{g2sns}
\end{align}
where $\langle \hat{n} \rangle = n \cosh(2 |\xi|) + \sinh^2{|\xi|}$. For small enough values of $|\xi|$, $g^{(2)}(0)$ can be less than unity, what characterizes a sub-Poissonian statistics. For $|\xi| \gg 1$, the second term in eq. (\ref{g2sns}) can be neglected and we have the following result,
\begin{equation}
	g^{(2)}(0) \approx 1 + \frac{2 (n^2 + n + 1)}{(2 n + 1)^2},
\end{equation}
which indicates super-Poissonian statistics.

\section*{III. THE MODEL}
\begin{figure}
	\includegraphics[height=2cm]{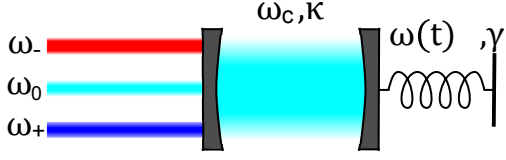}
	\caption{An optomechanical system in which the optical mode is pumped by three coherent radiation fields with frequency $\omega_+$, $\omega_-$ and $\omega_c$, and the mechanical mode is parametrically driven.}
	\label{fig1}	
\end{figure}

We consider an optomechanical system having a purely quadratic coupling between an optical mode of frequency $\omega_c$ and a mechanical parametric oscillator, whose frequency is given by $ \omega^2(t) = \omega_m^2 \bigl[1 + \epsilon \cos(\Omega t)\bigr]$, with $0 \le |\epsilon| < 1$ .The Hamiltonian of this system is given by ($\hbar = 1$),
\begin{multline}
	\hat{H} = \omega_c \hat{A}^{\dagger} \hat{A} + \frac{\omega_m}{2} \hat{P}^2 + \frac{\omega_m}{2} \bigl[1 + \epsilon \cos(\Omega t)\bigr] \hat{Q}^2 + g \hat{A}^{\dagger} \hat{A} \hat{Q}^2 \\+ \hat{H}_{dr},
	\label{h1}
\end{multline} 
where
\begin{equation}
	\hat{H}_{dr} = i E_+ \hat{A}^{\dagger} e^{-i \omega_+ t} + i E_0 \hat{A}^{\dagger} e^{-i \omega_c t} + i E_- \hat{A}^{\dagger} e^{-i \omega_- t} + h.c. 
	\label{hdr}
\end{equation}
$\hat{A}$ is the annihilation operator of the optical mode, $\hat{Q}$ and $\hat{P}$ are the dimensionless position and momentum operators of the mechanical mode, and $g$ is the optomechanical coupling parameter; $\omega_j$ and $E_j$($j=\pm,0$) are the frequency and the amplitude of the j'th coherent field. We assume that the optical field is strongly pumped and that $\omega_{\pm} = \omega_c \pm \Delta$. The value of $\Delta$ will be determined later in order to engineer the desired interaction between the optical and the mechanical modes.

Let us suppose that both the optical mode and the mechanical mode interact with the environment. In this situation, the dynamics of the system can be studied using the quantum Langevin equations (QLE) \cite{gardiner}:
\begin{align}
	\frac{d}{dt} \hat{A} =& -(\kappa + i \omega_c) \hat{A} - i g \hat{A} \hat{Q}^2 + E_+ e^{-i \omega_+ t} + E_0 e^{-i \omega_c t} \nonumber\\
	& \qquad \qquad \qquad \qquad + E_- e^{-i \omega_- t} + \sqrt{\kappa} \hat{A}_{in}, \label{qle1}\\
	\frac{d}{dt} \hat{Q} =& \omega_m \hat{P},\\
	\frac{d}{dt} \hat{P} =& -\omega_m \bigl[1 + \epsilon \cos(\Omega t)\bigr] \hat{Q} - 2 g \hat{A}^{\dagger} \hat{A} \hat{Q} - \gamma \hat{P} + \hat{\chi},
	\label{qle2}
\end{align}
where $\kappa$ and $\gamma$ are the  decay rate of the optical and mechanical modes, respectively. The last two equations can be decoupled, resulting in the following set of equations,
\begin{align}
	\frac{d}{dt} \hat{A} =& -(\kappa + i \omega_c) \hat{A} - i g \hat{A} \hat{Q}^2 + E_+ e^{-i \omega_+ t} + E_0 e^{-i \omega_c t}  \nonumber\\
	& \qquad \qquad \qquad \qquad + E_- e^{-i \omega_- t} + \sqrt{\kappa} \hat{A}_{in}, \label{qle1}\\
	\frac{d^2}{dt^2} \hat{Q} =& -\omega_m^2 \bigl[1 + \epsilon \cos(\Omega t) \bigr] \hat{Q} - 2 g \omega_m \hat{A}^{\dagger} \hat{A} \hat{Q} - \gamma \frac{d}{dt}\hat{Q} \nonumber\\
	& \qquad \qquad \qquad \qquad \qquad \qquad \qquad \qquad + \hat{\chi}.
	\label{qle2}
\end{align}
The operators $\hat{A}_{in}$ and $\hat{\chi}$ are (statistically independent) Gaussian noise operators whose first order correlation functions are equal to zero and 
the second order correlation functions are \cite{gardiner},
\begin{align}
	&\langle \hat{A}_{in}(t) \hat{A}_{in}^{\dagger}(t^{\prime}) \rangle = (\bar{n}_c + 1)\delta(t - t^{\prime}),\\
	&\langle \hat{A}_{in}^{\dagger}(t) \hat{A}_{in}(t^{\prime}) \rangle = \bar{n}_c \delta(t - t^{\prime}), \\
	&\langle \{ \hat{\chi}(t), \hat{\chi}(t^{\prime}) \} \rangle = \frac{2 \gamma}{\pi \omega_m} \int d \omega \omega \coth\left(\frac{\hbar \omega}{k_B T}\right) \cos[\omega (t - t^{\prime})], \label{cxi}
\end{align}
where $\bar{n}_c = [\exp(\hbar \omega_c/k_B T) - 1]^{-1}$ is the mean thermal excitation number of the optical environment, $T$ is the temperature, and $\{,\}$ denotes the anticommutator. We assume in this work that the optical frequency is high enough so that we can neglect $\bar{n}_c$. %If the mechanical oscillator has a good quality factor, $\omega_m \gg \gamma$, eq. (\ref{cxi}) can be approximated by, 
%\begin{equation}
%	\langle \{ \hat{\xi}(t) \hat{\xi}^{\dagger}(t^{\prime}) \} \rangle = 2 \gamma \coth\left(\frac{\hbar \omega}{k_B T}\right) \delta(t - t^{\prime}),
%\end{equation}
%what gives us a Markovian dynamics. 

In the strong pumping regime we can decompose the operators $\hat{A}$, $\hat{Q}$ and $\hat{P}$ as the sum of their expectation values and small fluctuations: $\hat{A} = \alpha + \hat{a}$, $\hat{Q} = \sigma + \hat{q}$ and $\hat{P} = \eta + \hat{p}$. In this regime we have the following equations for the expectation values $\alpha$ and $\sigma$,
\begin{align}
	\frac{d}{dt} \alpha =& -(\kappa + i \omega_c) \alpha - i g \alpha \sigma^2 + E_+ e^{-i \omega_+ t} + E_0 e^{-i \omega_c t} \nonumber \\
	& \qquad \qquad \qquad \qquad \qquad \qquad + E_- e^{-i \omega_- t}, \label{alpha}\\	
	\frac{d^2}{dt^2} \sigma =& -\omega_m^2 \bigl[1 + \epsilon \cos(\Omega t)\bigr] \sigma - 2 g \omega_m |\alpha|^2 \sigma - \gamma \frac{d}{dt}\sigma.
	\label{sigma}
\end{align}
Eqs. (\ref{alpha}) and (\ref{sigma}) are nonlinear differential equations and their solution can be very hard to find. However, given that in the majority of the experimental realizations of an optomechanical system with quadratic coupling the coupling parameter $g$ is very small compared to $\kappa$ and $\omega_m$, we may try a perturbative approach. In order to do that we have to make an additional assumption about the asymptotic value of $\sigma(t)$, or: $\lim_{t \to \infty} \sigma(t) \neq \infty$. This last condition is not trivially satisfied, as we are dealing with a parametric oscillator, and it must be verified at the end of the procedure. If this last condition is satisfied, then, for a sufficiently small $g$, we may find a perturbative solution. Let us call $\alpha_0$ the unperturbed solution of eq. (\ref{alpha}), i.e. neglecting the optomechanical interaction term. Thus, the asymptotic solution of eq. (\ref{alpha}) for $g = 0$ can be easily found and is given by,
\begin{equation}
	\alpha_0(t) = \frac{E_+}{\kappa + i \Delta} e^{-i \omega_+ t} + \frac{E_0}{\kappa} e^{-i \omega_c t} + \frac{E_-}{\kappa - i \Delta} e^{-i \omega_- t},
	\label{solalpha0}
\end{equation}
where $E_0/\kappa$ and $E_{\pm}/(\kappa \pm i \Delta)$ are assumed to be real, what can be done by adjusting the phase of $E_j$. Replacing $\alpha_0(t)$ in eq. (\ref{sigma}) gives us the following equation,
\begin{multline}
	\frac{d^2}{dt^2} \sigma = -\omega_m^2 \bigl[1 + 2 \delta + \epsilon \cos(\Omega t)  + 2 \lambda \cos(\Delta t) \\+ 2 \mu \cos(2 \Delta t) \bigr] \sigma - \gamma \frac{d}{dt}\sigma,
	\label{sigma0}
\end{multline}
where 
\begin{align}
	\delta =& \frac{g}{\omega_m} \left( \frac{|E_+|^2}{\kappa^2 + \Delta^2} + \frac{|E_0|^2}{\kappa^2} + \frac{|E_-|^2}{\kappa^2 + \Delta^2}\right),\\
	\lambda =& \frac{2 g}{\omega_m} \frac{|E_0|}{\kappa} \left( \frac{|E_+| + |E_-|}{\sqrt{\kappa^2 + \Delta^2}} \right),\\
	\mu =& \frac{2 g}{\omega_m} \frac{|E_+ E_-|}{\kappa^2 + \Delta^2}.
\end{align}
Let us analyze some important facts about eq. (\ref{sigma0}): firstly, it will give us a corrected solution, as the optomechanical interaction term is taken into account. Secondly, the optomechanical interaction gives rise to an effective frequency $\tilde{\omega}_m$ for the mechanical oscillation, defined as $\tilde{\omega}_m = \omega_m \sqrt{1 + 2 \delta}$. For reasons that will become clear later, we will choose $\Omega = \Delta = 2 \tilde{\omega}_m$. This choice has very important consequences, as the presence of terms proportional to $\cos(2 \tilde{\omega}_m t)$ has the potential to lead to parametric instability. To avoid this scenario, we impose that $\epsilon = - 2 \lambda$. Given that the presence of a term proportional to $\cos(4 \tilde{\omega}_m t)$ does not give rise to instability issues (except for large $\epsilon$, a situation that we will not consider here), it can be shown that, if the last condition is fulfilled, the asymptotic solution of eq. (\ref{sigma0}) is $\sigma(t) = 0$ (a discussion about instability issues can be found in the Appendix). This result agrees with our initial assumptions. Now, if we try to obtain a more accurate solution for $\alpha(t)$ by replacing the solution found for eq. (\ref{sigma0}) in eq. (\ref{alpha}) and do the whole procedure again (i.e., use this corrected solution of $\alpha(t)$ to obtain a corrected solution of $\sigma(t)$), we will find the same results. Therefore, we have that the asymptotic value of $\alpha(t)$ is given by eq. (\ref{solalpha0}) and the asymptotic value of $\sigma(t)$ is zero. Consequently, the asymptotic value of $\eta(t)$ is also zero.

Substituting those results in Hamiltonian (\ref{h1}),
\begin{multline}
	\hat{H} = \omega_c \hat{a}^{\dagger} \hat{a} + \frac{\omega_m}{2} \hat{p}^2 + \frac{\omega_m}{2} \bigl[1 + \epsilon \cos(2 \tilde{\omega}_m t)\bigr] \hat{q}^2 + g \bigl[|\alpha(t)|^2 \\+ \alpha(t) \hat{a}^{\dagger} + \alpha^{\ast}(t) \hat{a} + \hat{a}^{\dagger} \hat{a}\bigr] \hat{q}^2.
	\label{h2}
\end{multline}
The last term in Hamiltonian (\ref{h2}), $g \hat{a}^{\dagger} \hat{a} \hat{q}^2$, is much smaller than the others terms, so it is going to be neglected. Using eq. (\ref{solalpha0}),
\begin{multline}
	\hat{H} = \omega_c \hat{a}^{\dagger} \hat{a} + \frac{\omega_m}{2} \hat{p}^2 + \frac{\omega_m}{2} \bigl[1 + 2 \delta \bigr] \hat{q}^2 + \bigl[\mu \cos(4 \tilde{\omega}_m t) \\+ g \alpha(t) \hat{a}^{\dagger} + g \alpha^{\ast}(t) \hat{a} \bigr] \hat{q}^2.
	\label{h3}
\end{multline}
Rewriting Hamiltonian (\ref{h3}) in terms of the annihilation operator of the mechanical oscillator with effective frequency $\tilde{\omega}_m$, $\hat{b}$, we have,
\begin{multline}
	\hat{H} = \omega_c \hat{a}^{\dagger} \hat{a} + \tilde{\omega}_m  \hat{b}^{\dagger} \hat{b} + \Bigl[ \frac{\omega_m^2}{2 \tilde{\omega}_m} \mu \cos(4 \tilde{\omega}_m t) + g^{\prime} \alpha(t) \hat{a}^{\dagger} \\+ g^{\prime} \alpha^{\ast}(t) \hat{a} \Bigr] (\hat{b} + \hat{b}^{\dagger})^2,
	\label{h3.1}
\end{multline}
where, $g^{\prime} = g \omega_m/(2 \tilde{\omega}_m)$. Going to the interaction picture we have the following Hamiltonian,
\begin{equation}
	\hat{H}_I = \hat{H}_R + \hat{H}_{NR}
	\label{h4}
\end{equation}
where the resonant part is
\begin{equation}
	\hat{H}_R = \hat{a}^{\dagger} \bigl(g_- \hat{b}^2 + 2 g_0 \hat{b}^{\dagger} \hat{b} + g_+ \hat{b}^{\dagger 2} + g_0\bigl) + h.c.,
	\label{hr}
\end{equation}
and the nonresonant part is
\begin{multline}
	\hat{H}_{NR} = \frac{\omega_m^2}{4 \tilde{\omega}_m} \mu \cos(4 \tilde{\omega}_m t) \bigl(\hat{b}e^{-i \tilde{\omega}_m t}  + \hat{b}^{\dagger} e^{i \tilde{\omega}_m t} \bigr)^2 \\+ \hat{a}^{\dagger} \bigl(2 \hat{b}^{\dagger} \hat{b} + 1 \bigr) \bigl(g_+ e^{-2 i \tilde{\omega}_m t} + g_-  e^{2 i \tilde{\omega}_m t} \bigl)\\ + \hat{a}^{\dagger} \hat{b}^2 \bigl(g_+ e^{-4 i \tilde{\omega}_m t} + g_0  e^{-2 i \tilde{\omega}_m t} \bigr) + \hat{a} \hat{b}^{\dagger 2} \bigl(g_- e^{4 i \tilde{\omega}_m t} \\+ g_0  e^{2 i \tilde{\omega}_m t} \bigr) + h.c.,
	\label{hnr}
\end{multline}
where $g_{0} = g^{\prime} E_j/\kappa$ and $g_{\pm} = g^{\prime} E_j/(\kappa \pm i \Delta)$. It is important to note that the values of $\Omega$ and $\Delta$ were decisive to determine which terms of Hamiltonian (\ref{h3.1}) are resonant and which are nonresonant.  

In principle, the open system dynamics of the optomechanical system can be studied using the QLE's for the fluctuation operators $\hat{a}$ and $\hat{b}$, which can be obtained using the simplified Hamiltonian (\ref{h4}). However, although the QLE's obtained with Hamiltonian (\ref{h4}) are simpler than the QLE's (\ref{qle1}) and (\ref{qle2}), they are still nonlinear QLE's, whose analytical solutions are difficult to find. Therefore, we are going to study the dynamics of our system using a master equation approach. With the simplified Hamiltonian (\ref{h4}), our system is described by the following master equation,
\begin{equation}
	\frac{d}{dt} \hat{\rho} = -i [\hat{H}_I, \hat{\rho}] + \kappa \hat{\mathcal{L}}[\hat{a}] \hat{\rho} + (\bar{n}_m + 1) \gamma \hat{\mathcal{L}}[\hat{b}] \hat{\rho} + \bar{n}_m \gamma \hat{\mathcal{L}}[\hat{b}^{\dagger}] \hat{\rho},
	\label{me1}
\end{equation}
where $\bar{n}_m = [\exp(\hbar \omega_m/k_B T) - 1]^{-1}$ is the mean thermal excitation number of the mechanical environment and $\hat{\mathcal{L}}[\hat{c}] \hat{\rho}= 2 \hat{c} \hat{\rho} \hat{c}^{\dagger} - \hat{c}^{\dagger} \hat{c} \hat{\rho} - \hat{\rho} \hat{c}^{\dagger} \hat{c}$, is the Lindbladian. As it is very hard to find analytical solutions of eq. (\ref{me1}), we have treated the problem numerically.

\section*{IV. DISSIPATION INDUCED GENERATION OF SQUEEZED NUMBER STATES}
We assume that our system operates in the deep resolved sideband limit ($\kappa \ll \omega_m$), what allows us to make the rotating wave approximation and neglect the nonresonant part of Hamiltonian (\ref{h4}), $\hat{H}_{NR}$. We also assume that $g_-/g_+ = \coth^2{r}$ and $g_-/g_0 = \coth{r}$. In this situation we can define the coupling parameter $\mathcal{G} = g_- - g_+$, and the Hamiltonian (\ref{h4}) can be written in terms of the Bogoliubov annihilation operator $\hat{\beta}(r) = \cosh(r) \hat{b} + \sinh(r) \hat{b}^{\dagger}$,
\begin{equation}
	\hat{H}_I = \mathcal{G} \bigl[\hat{a}^{\dagger} \hat{\beta}^2(r) + \hat{a} \hat{\beta}^{\dagger 2}(r) \bigr].
	\label{h5}
\end{equation} 
This Hamiltonian allows the creation (annihilation) of one photon and annihilation (creation) of two excitations of the Bogoliubov mode. If we suppose, further, that the mechanical damping can be neglected, or $\gamma = 0$, the open system dynamics allows the optical mode to cool the Bogoliubov mode, leading the system to the states $|\psi_0 \rangle = |0 \rangle_a |\xi, 0 \rangle_b$ or $|\psi_1 \rangle = |0 \rangle_a |\xi, 1 \rangle_b$. One way to conclude this is to observe that states $|\psi_0 \rangle$ and $|\psi_1 \rangle$ are dark states of the dynamics, i.e.
\begin{align}
	\hat{H} | \psi_n\rangle &= 0,\\
	\hat{\mathcal{L}}[\hat{a}] (|\psi_n \rangle \langle \psi_n|) &= 0.
\end{align} 
Indeed, if we suppose that in the steady state the cavity must be in the vacuum state, then the states $|\psi_0 \rangle$ and $|\psi_1 \rangle$ are the only (orthogonal) dark states of the system. This means that any coherent or incoherent mixture of the states $|\psi_0 \rangle$ and $|\psi_1 \rangle$ is a steady state of the system. The exact form of the steady state of the system is determined by the initial state of the system, in particular, if the initial state of the mechanical oscillator is a superposition of even (odd) number states, the steady state will be $|\psi_0 \rangle = |0 \rangle_a |\xi = r, 0 \rangle_b$ ($|\psi_1 \rangle = |0 \rangle_a |\xi = r, 1 \rangle_b$). Thus, in the absence of mechanical damping, $r$ indeed corresponds to the squeezing parameter of the mechanical state.

In the situation in which $\kappa \gg \mathcal{G}$ and $\kappa \gg \gamma$, we may proceed with the adiabatic elimination of the optical mode. By standard methods \cite{gardiner, carmichael}, we obtain the following master equation,
\begin{equation}
	\frac{d}{dt} \hat{\rho}_b = \left[ (\bar{n}_m + 1) \gamma \hat{\mathcal{L}}[\hat{b}] + \bar{n}_m \gamma \hat{\mathcal{L}}[\hat{b}^{\dagger}] + \frac{\mathcal{G}^2}{\kappa} \hat{\mathcal{L}}[\hat{\beta}^2(r)] \right] \hat{\rho}_b,
	\label{meae}
\end{equation}
where $\hat{\rho}_b = Tr_a[\hat{\rho}]$. From eq. (\ref{meae}) we note that the the optical field acts as an engineered reservoir for the mechanical oscillator. It is clear that in the presence of mechanical damping ($\gamma \neq 0$) the states $|\psi_0 \rangle$ and $|\psi_1 \rangle$ are no longer dark states of the dynamics. 

It is important to stress that the possibility of engineering a squeezed vacuum state and a squeezed one phonon state strongly depends on the validity of the rotating wave approximation. If the nonresonant significantly contribute to the dynamics, what is expected to happen if our system does not operate in the deep resolved sideband regime, then a nonresonant heating of the Bogoliubov mode would occur. This certainly will cause deleterious effects on the generation of the target states.

\section*{V. RESULTS}
In this section we show the results obtained by numerically solving eq. (\ref{me1}). We calculate the temporal evolution of the system, and directly find its steady state. The only drawback of this approach is that for larger values of $r$ it is necessary to consider a Hilbert space with larger dimensions. This severely limits the values of $r$ considered here. Another important point is that, although it would be more suitable from a theoretical point of view to scale the time in units of $\kappa/\mathcal{G}^2$ (as $\kappa/\mathcal{G}^2$ determines a time scale of the system, at least in situations where the optical mode can be adiabatic eliminated, which is the most common situation in the laboratory), we choose to scale the time in units of $\kappa^{-1}$. The reason is because scaling the time in units of $\kappa/\mathcal{G}^2$ would require to keep $\mathcal{G}$ constant, what means that (for larger values of $r$) the effective couplings $g_i$, specially $g_-$, would assume values that are orders of magnitude greater than the values that can be obtained experimentally with the current technology. In all calculations we use $g_- = 0.01 \kappa$, $g_0 = 0.01 \kappa \tanh r$ and $g_+ = 0.01 \kappa \tanh^2 r$. The numerical calculations were done using the software QuTiP \cite{johansson2}.
\begin{figure}
	\centering	
	\subfigure{
		\includegraphics[width=4.1cm]{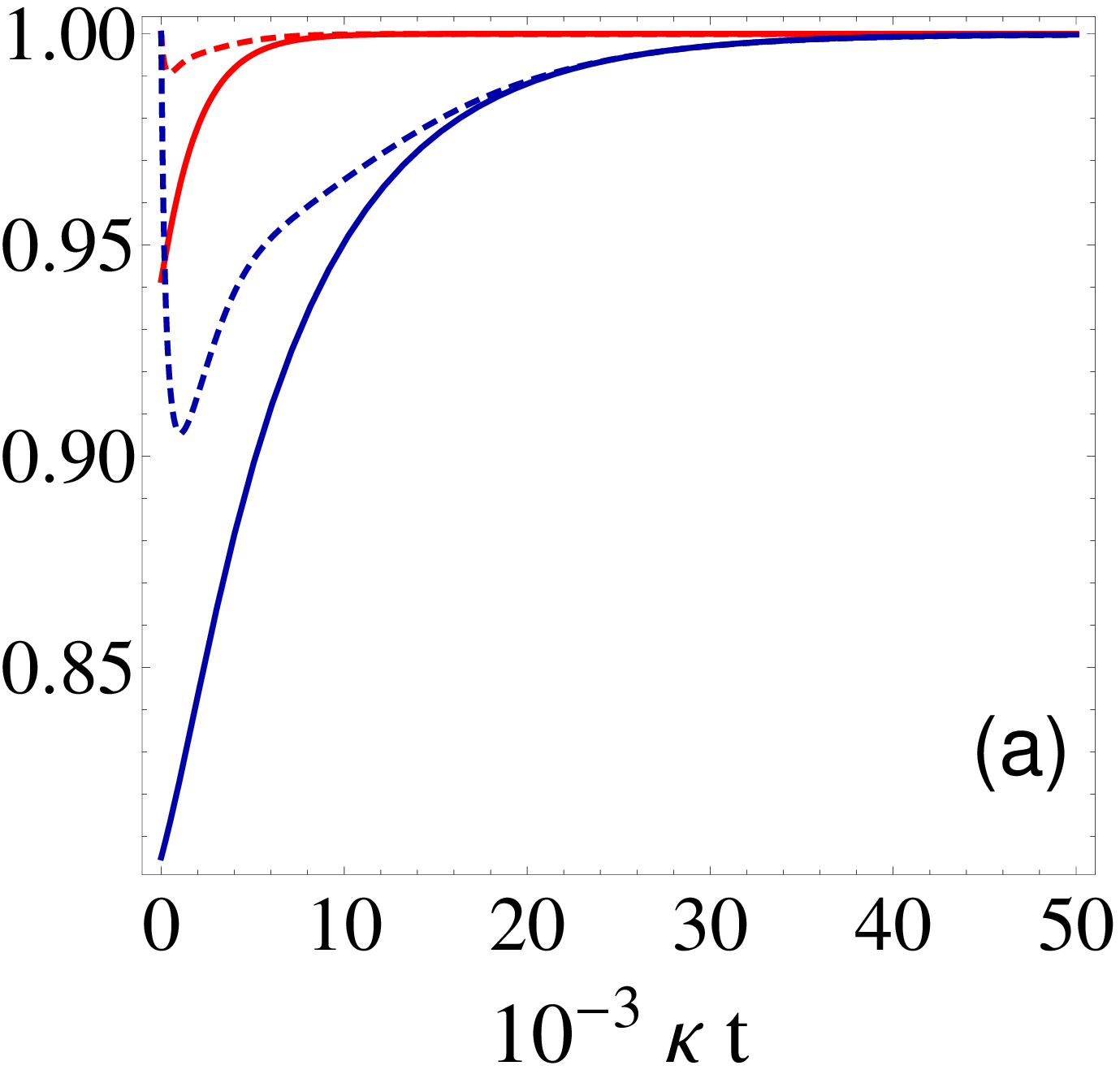}
		\label{fig2a}
	}\hspace{-0.2cm}
	\subfigure{
		\includegraphics[width=4cm]{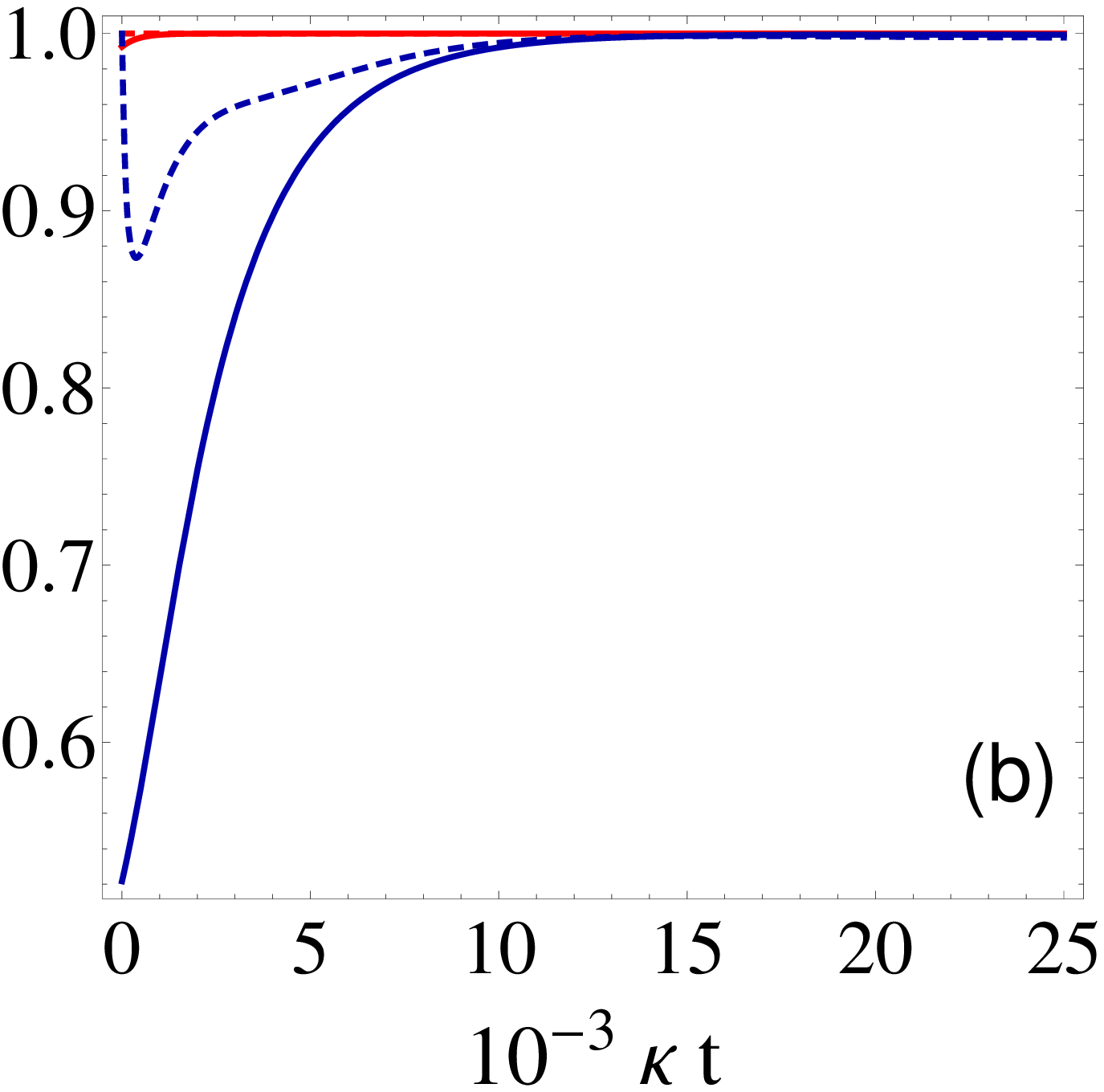}
		\label{fig2b}
	}
	\caption{Fidelity (continuous curves) and purity (dashed curves) as a function of $\kappa t$ for $\gamma = 0$. (a) $r = 0.5$ in the red curves and $r = 1$ in the blue curves. The initial state in both curves is $ |n_a = 0, n_b = 0 \rangle$. (b) $r = 0.1$ in the red curves and $r = 1$ in the blue curves. The initial state in both curves is $|n_a = 0, n_b = 1 \rangle$.}
	\label{fig2}	
\end{figure}

In figure \ref{fig2} we plot the fidelity $F = \sqrt{\langle \psi | \hat{\rho} | \psi \rangle}$ between the state of the system and the target state $|\psi \rangle$, and the purity of the state of the system in the absence of mechanical damping. The target state is the state $|0 \rangle_a |\xi = r, 0 \rangle_b$ in (a) and the state $|0 \rangle_a |\xi = r, 1 \rangle_b$ in (b). It is clear that after some time the state of the system evolves to the corresponding target states. Using eq. (\ref{ux1}), we can calculate the amount of squeezing reached in figure \ref{fig2a}, which is approximately $8.69$ $dB$ in the blue curve and $4.34$ $dB$ in the red curve. In figure \ref{fig2b} the amount of squeezing can be calculated using eq. (\ref{ux1sn}) for $n=1$. In the red curve, given that $r<\ln \sqrt{3}$, no squeezing is observed. However, for small values of $r$, we obtain $g^{(2)}(0) = 0.058$, i.e., a sub-Poissonian statistics emerges (see the red curve). In the blue curve we have $r = 1$; the system now presents a super-Poissonian statistics, and the squeezing is approximately $3.91$ $dB$. It is interesting to note that if we prepare the mechanical oscillator in an initial one phonon state and set $r = 0$ (what means that $E_0 = E_+ = 0$), then the mechanical oscillator will remain in the one phonon state. This is a direct consequence of the quadratic coupling and the assumption that $\gamma = 0$. 
\begin{figure}
	\centering	
	\subfigure{
		\includegraphics[width=4.1cm]{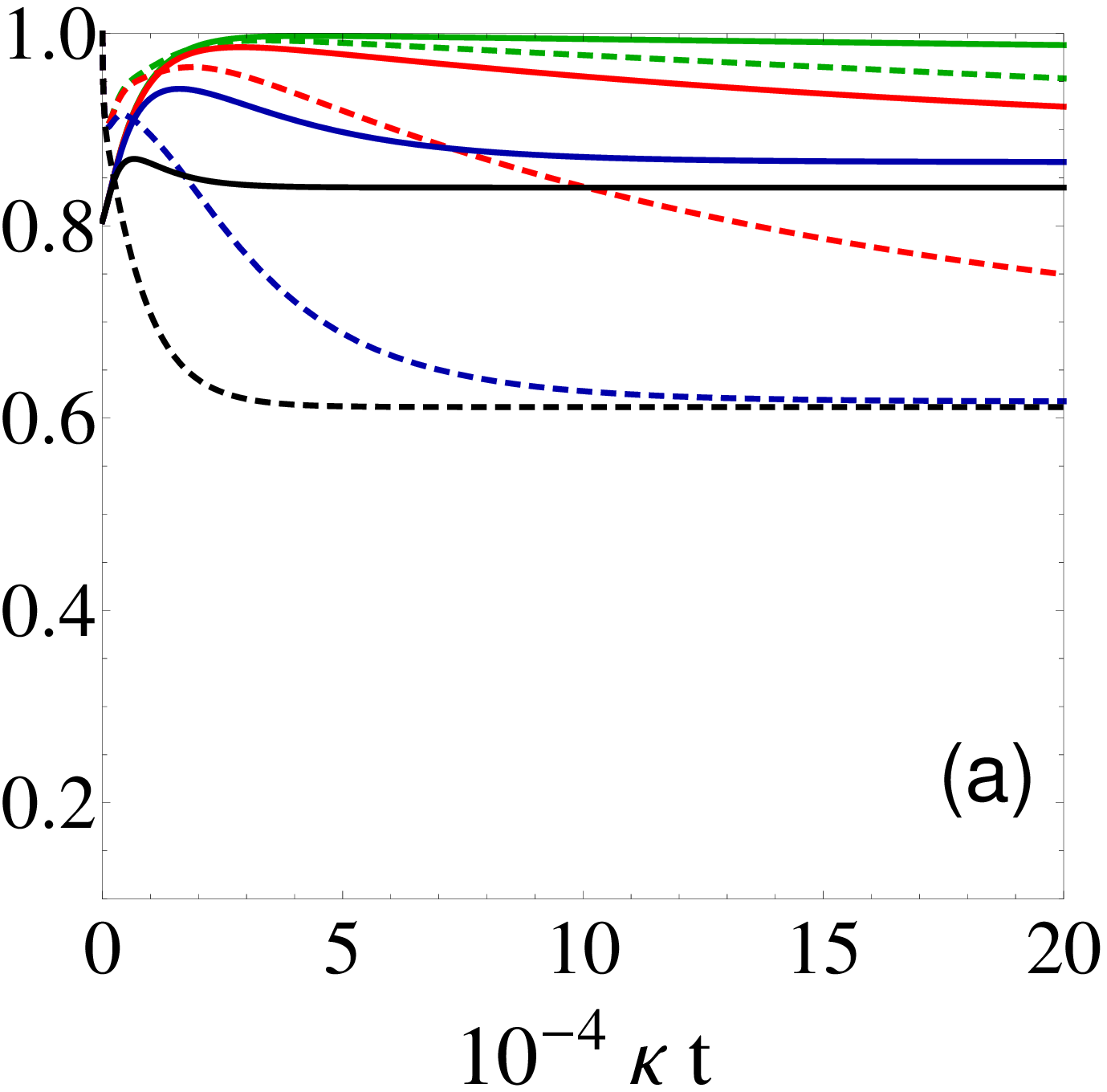}
		\label{fig3a}
	}\hspace{-0.2cm}
	\subfigure{
		\includegraphics[width=4.1cm]{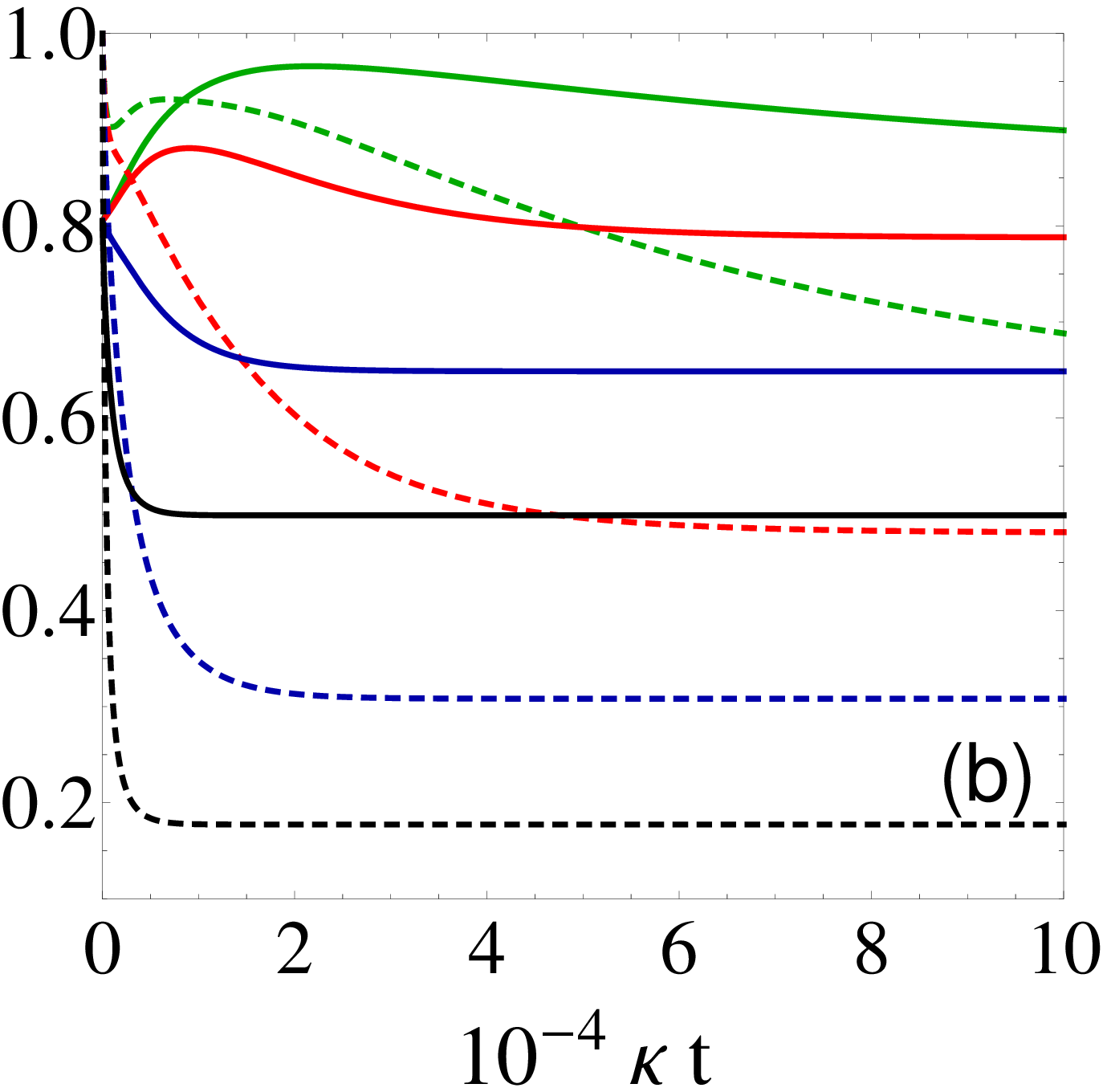}
		\label{fig3b}
	}
	\subfigure{
		\includegraphics[width=4.1cm]{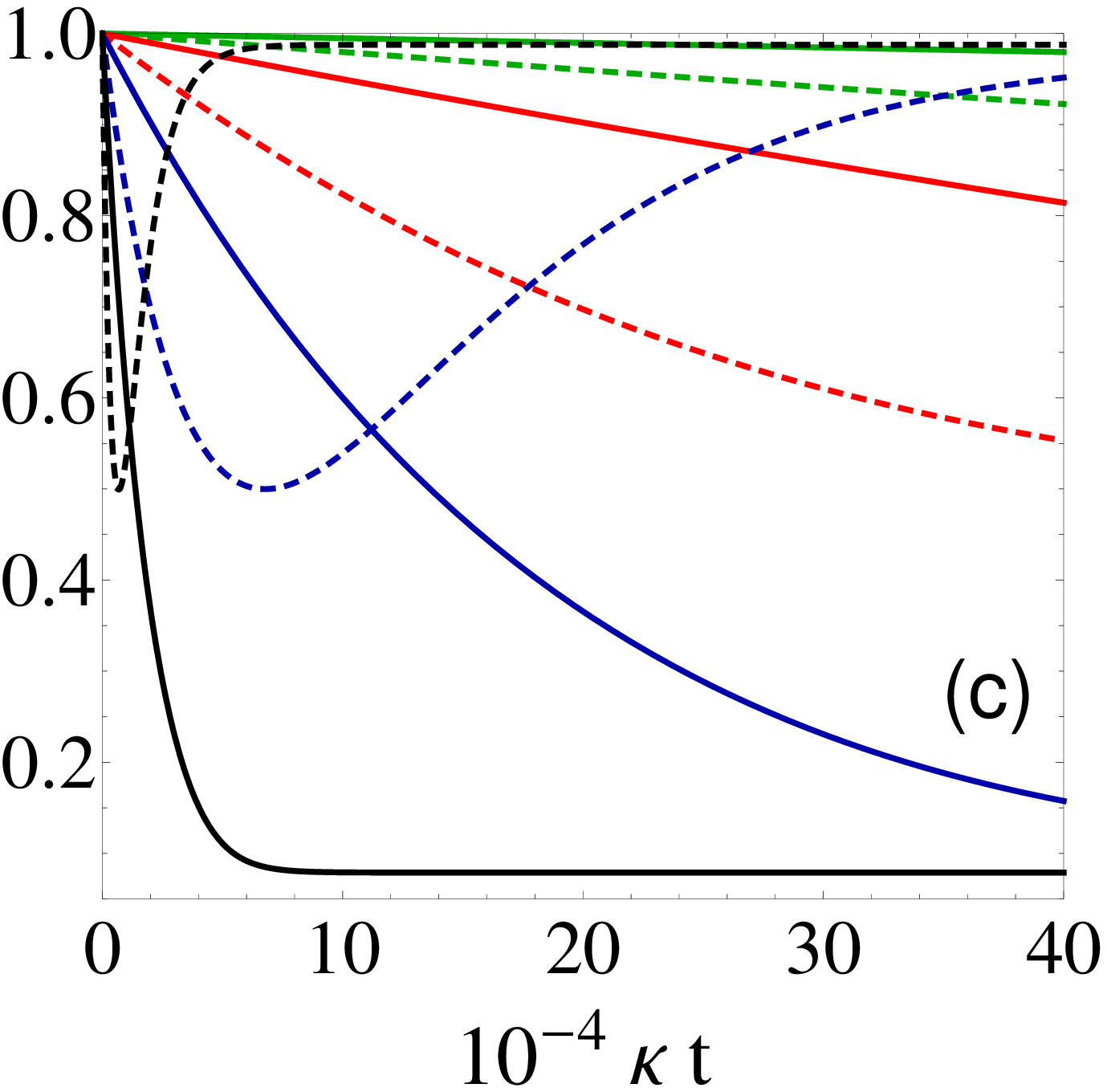}
		\label{fig3c}
	}\hspace{-0.2cm}
	\subfigure{
		\includegraphics[width=4.1cm]{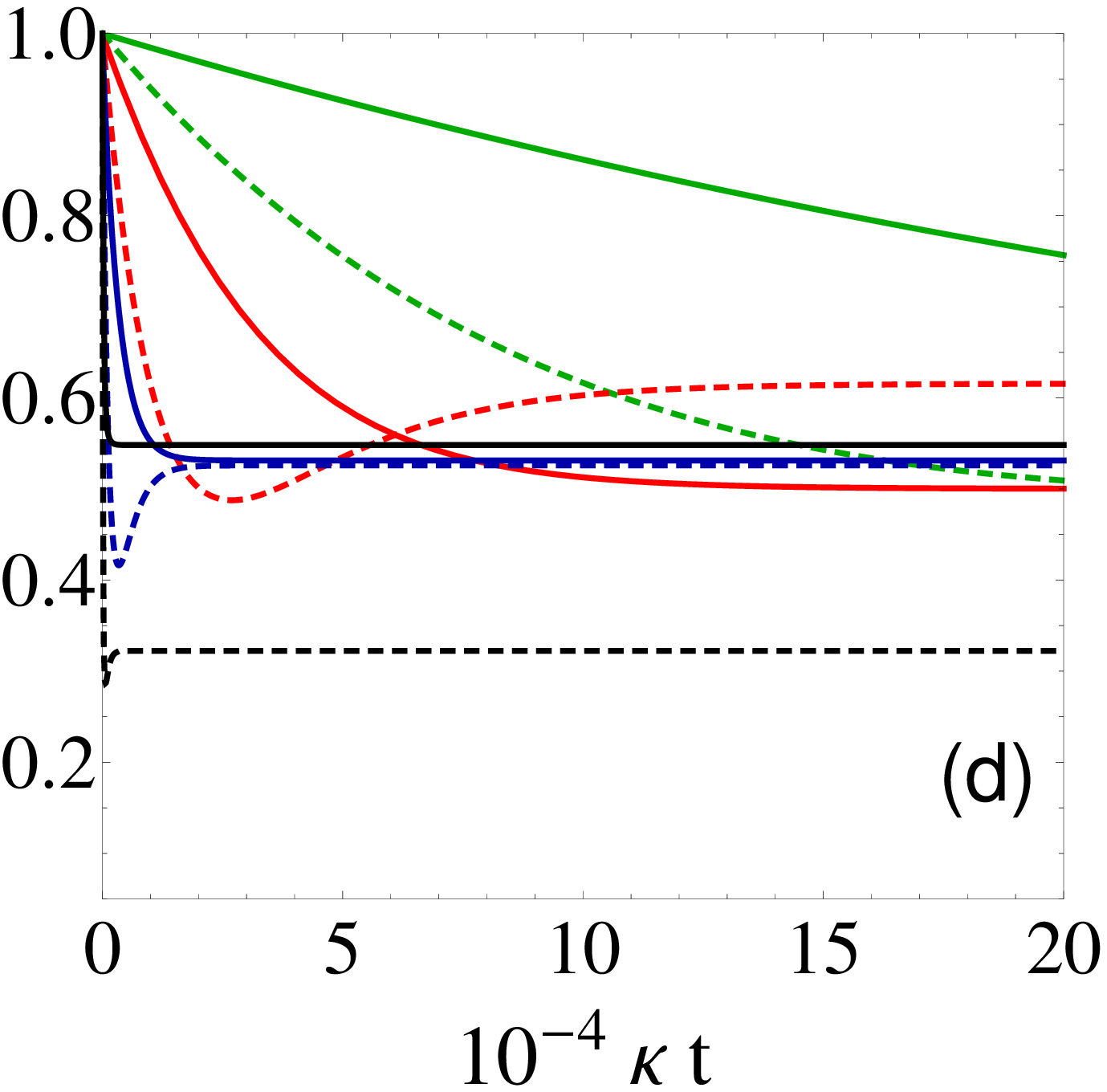}
		\label{fig3d}
	}
	\caption{Fidelity (continuous curves) and purity (dashed curves) as a function of $\kappa t$. (a) $r = 1$, $\bar{n}_m = 0$ and the initial state is $ |n_a = 0, n_b = 0 \rangle$. (b) $r = 1$, $\bar{n}_m = 10$ and the initial state is $ |n_a = 0, n_b = 0 \rangle$.(c) $r = 0.1$, $\bar{n}_m = 0$ and the initial state is $|n_a = 0, n_b = 1 \rangle$. (d) $r = 0.1$, $\bar{n}_m = 10$ and the initial state is $|n_a = 0, n_b = 1 \rangle$. In the black curves $\gamma = 10^{-4} \kappa$, in the blue curves $\gamma = 10^{-5} \kappa$, in the red curves $\gamma = 10^{-6} \kappa$ and in the green curves $\gamma = 10^{-7} \kappa$.}
	\label{fig3}	
\end{figure}

In figure \ref{fig3} we plot again the fidelity $F$ between the state of the system and the target state $|\psi \rangle$, as well as the purity of the state of the system. The target state is the state $|0 \rangle_a |\xi = 1, 0 \rangle_b$ in \ref{fig3a} and \ref{fig3b}, and the state $|0 \rangle_a |\xi = 0.1, 1 \rangle_b$ in \ref{fig3c} and \ref{fig3d}. Firstly, for the parameters' values used, the adiabatic approximation holds, so that the mechanical oscillator is effectively coupled to an engineered reservoir with a decay rate $\mathcal{G}^2/\kappa = 10^{-4} \sech^4(r) \kappa$. So, we expect our results to show the existence of two time scales; the first one determined by the coupling to the engineered reservoir and the second one determined by the coupling to the `real' mechanical reservoir. The results in figure \ref{fig3} show exactly the expected behavior. In the black curves we have $\gamma = 10^{-4} \kappa$ and $\bar{n} = 0 ,10$, what implies that coupling to the real reservoir is always equal or greater than the coupling to the effective reservoir. Consequently, the state of the system never reaches the target states. However, in the green curves we have $\gamma = 10^{-7} \kappa$ and $\bar{n} = 0 ,10$, or the coupling to the engineered reservoir greater than the coupling to the `real' reservoir, at least for the values of $r$ used. In this case, the results clearly show the existence of the two time scales and, at least for $\bar{n} = 0$, the state of the system presented a high fidelity relatively to the target states for a range of values of $\kappa t$. As expected, the state of the system is very well approximated by the target state for values of $t$ such that $t \gg \kappa/\mathcal{G}^2$ and $t \ll [(\bar{n}_m + 1) \gamma]^{-1}$. The only counter-intuitive behavior noted in figure \ref{fig3} is that if we compare the black continuous curves in figures \ref{fig3c} and \ref{fig3d}, we note that the fidelity $F$ seems to converge to a larger value for $\bar{n}_m = 10$ than for $\bar{n}_m = 0$. Indeed, the same behavior occurs for the other continuous curves in figures \ref{fig3c} and \ref{fig3d}, and it will be explained later when we consider the steady states of the system. It is important to point that the simulations of the full master equation, containing both optical and mechanical damping, show just one steady state, what means that independently of the initial state being an even or an odd Fock state (or any other state), the steady state of the system will be the same. The question now turns to be whether this steady state will show quantum features like squeezing and sub-Poissonian statistics.

\begin{figure}
	\centering	
	\subfigure{
		\includegraphics[width=4.2cm]{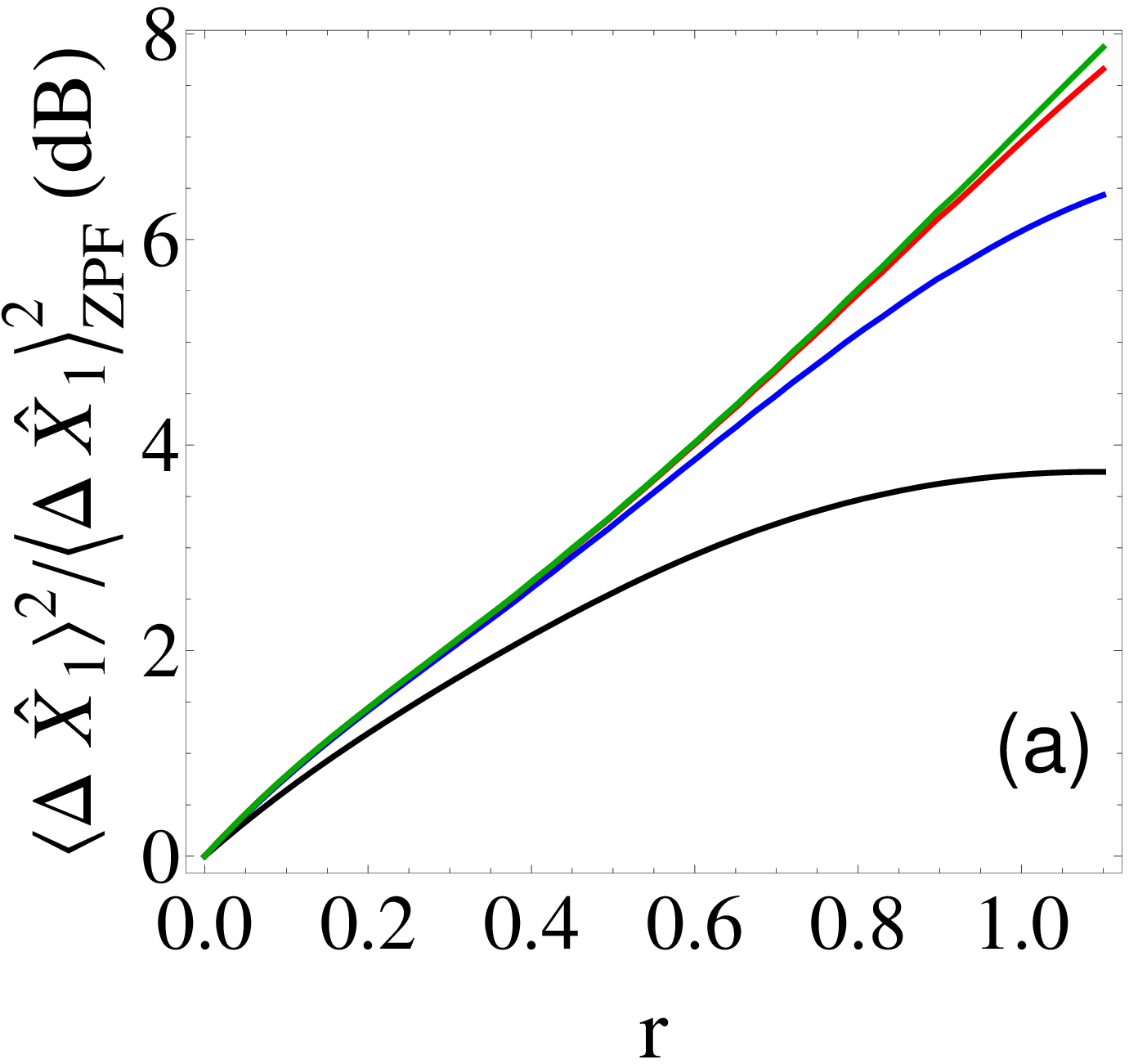}
		\label{fig4a}
	}\hspace{-0.3cm}
	\subfigure{
		\includegraphics[width=4.2cm]{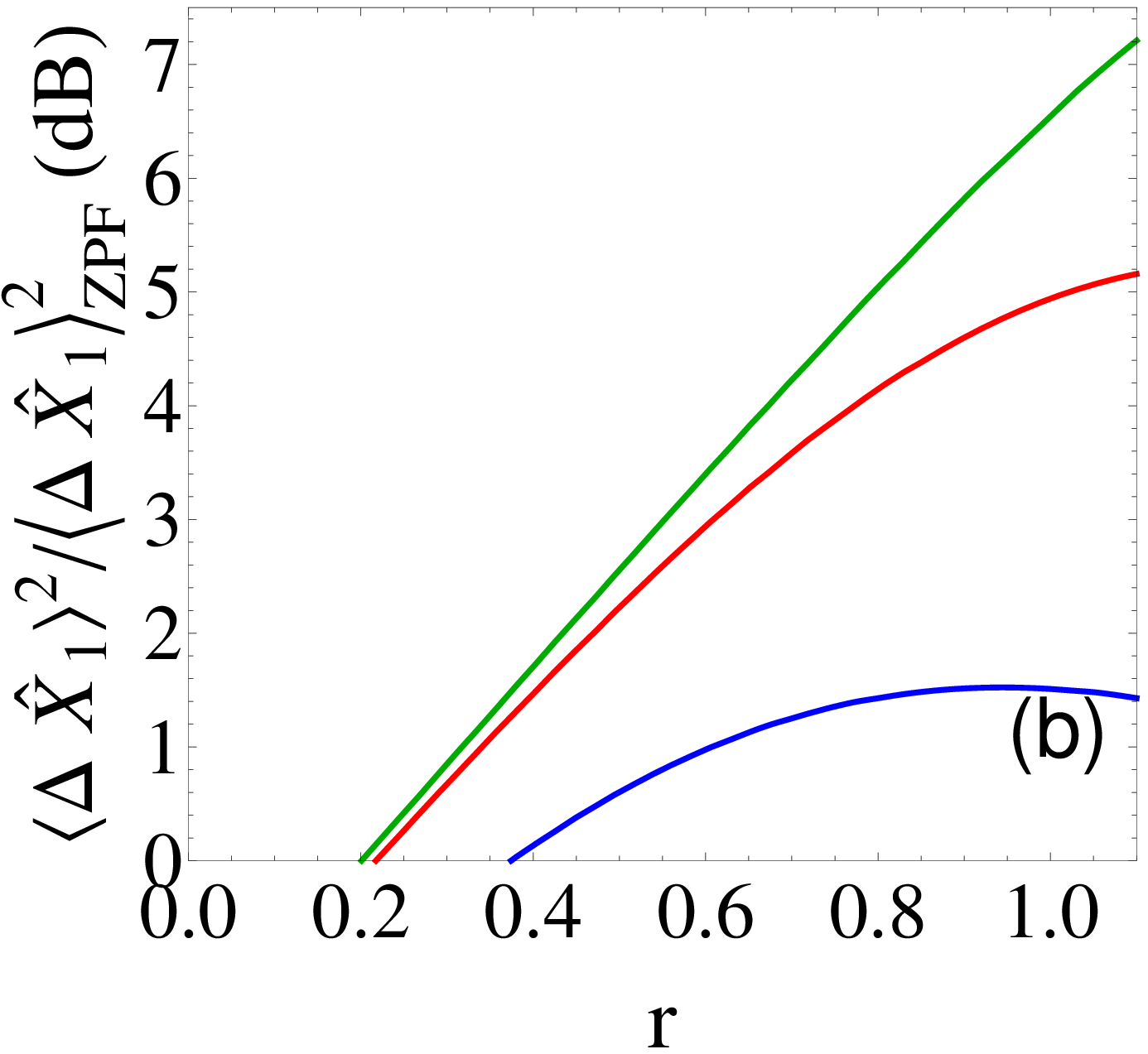}
		\label{fig4b}
	}
	\caption{Quadrature squeezing of $\hat{X}_1$ against $r$ in the steady state. In the black curves $\gamma = 10^{-4} \kappa$, in the blue curves $\gamma = 10^{-5} \kappa$, in red curves $\gamma = 10^{-6} \kappa$ and in the green curves $\gamma = 10^{-7} \kappa$. In (a) $\bar{n}_m = 0$ and in (b) $\bar{n}_m = 10$.}
	\label{fig4}	
\end{figure}
To answer this question, in figure \ref{fig4} we have the quadrature squeezing of $\hat{X}_1$ for the steady state of the field for different values of $\gamma$ and $\bar{n}_m$. As expected, the existence of mechanical damping has deleterious effects over the generation of squeezed states, the degree of squeezing being smaller for larger values of $\gamma$ and $\bar{n}$, however, it is still possible to reach squeezing above $7.9$ $dB$. We believe that for larger values of $r$ it is possible to reach even larger degrees of squeezing, although we could not simulate it because of the large size of the Hilbert space involved. An interesting feature can be observed in the black curve of figure \ref{fig4a} and in the blue curve of figure \ref{fig4b}, the degree of squeezing starts to decrease for larger values of $r$. This phenomenon was observed by Kronwald et al. \cite{kronwald}, and has to do with the fact that for larger values of $r$, the coupling parameter $\mathcal{G}$ is very small; indeed $\mathcal{G} \to 0$ for $r \to \infty$. Consequently, there is an optimal value $r_{opt}$ above which the coupling of the optical mode to the Bogoliubov mode is weak compared to the decoherence rate of the mechanical mode, and the generation of squeezing is compromised. Kronwald et al. also have found that for smaller values of $\gamma$ and $\bar{n}$, we have larger values for $r_{opt}$. This explains why we could not observe $r_{opt}$ in the others curves. It is important to stress that in the presence of mechanical damping, $r$ does not correspond to the squeezing parameter of the mechanical state anymore. This makes clear the existence of an optimal value for $r$. 

\begin{figure}
	\centering	
	\subfigure{
		\includegraphics[width=4.25cm]{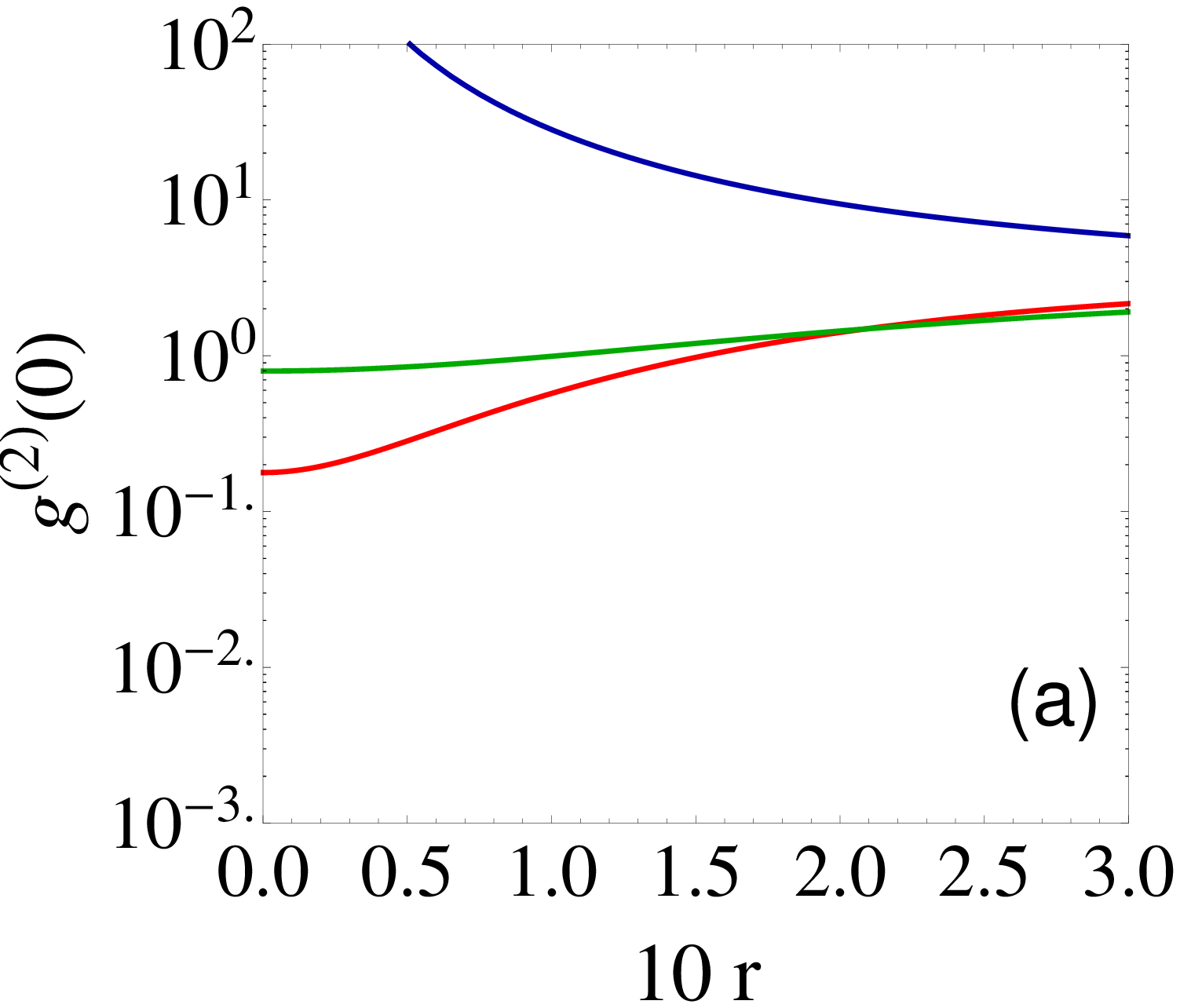}
		\label{fig5a}
	}\hspace{-0.4cm}
	\subfigure{
		\includegraphics[width=4.25cm]{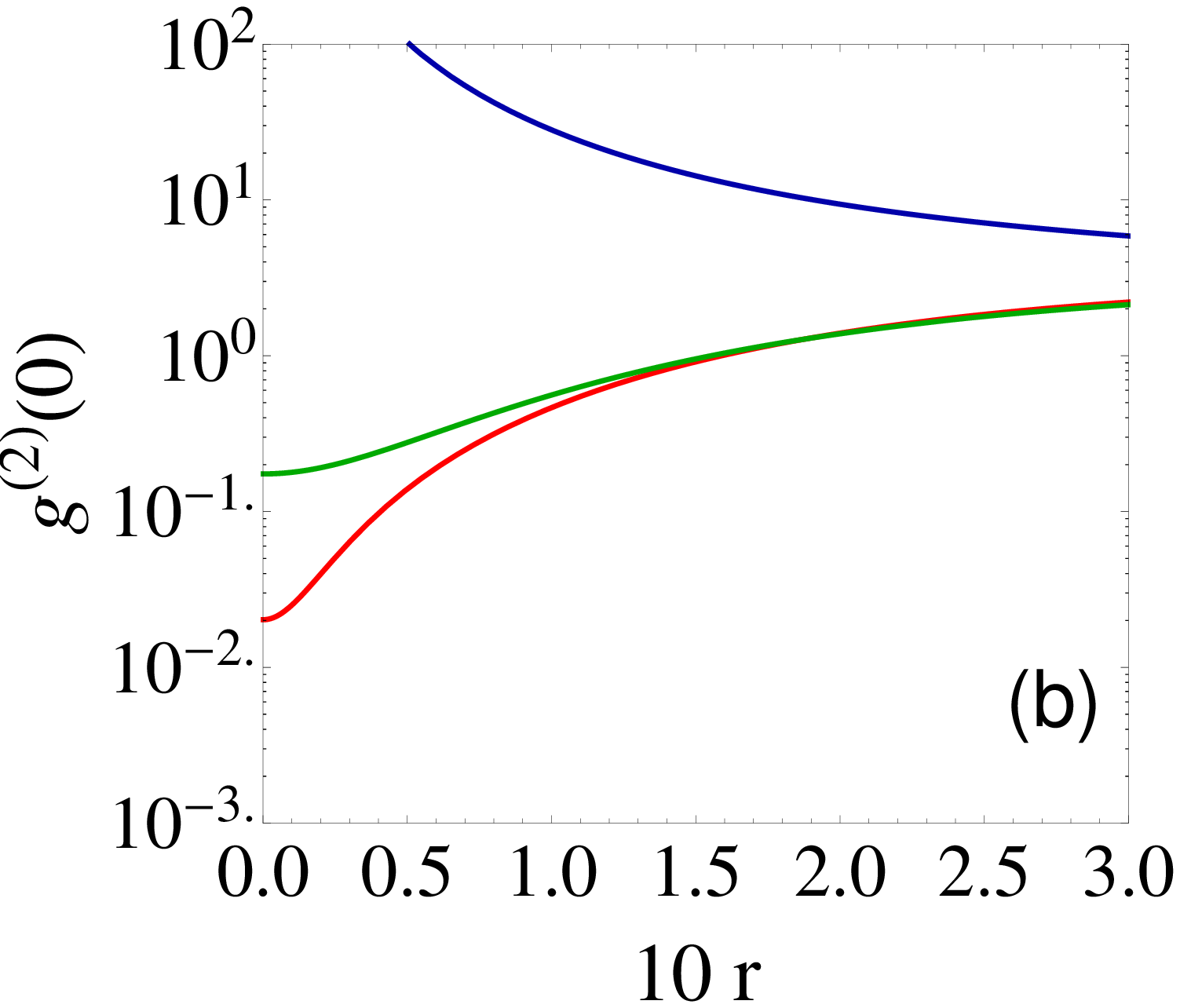}
		\label{fig5b}
	}
	\caption{$g^{(2)}(0)$ of the mechanical oscillator against $r$ in the steady state. In the blue curves $\bar{n}_m = 0$, in the red curves $\bar{n}_m = 10$ and in the green curves $\bar{n}_m = 100$. In (a) $\gamma = 10^{-6} \kappa$ and in (b) $\gamma = 10^{-7} \kappa$.}
	\label{fig5}	
\end{figure}
There is still the question of whether this system presents sub-Poissonian statistics in the presence of mechanical damping. In figure \ref{fig5} we have $g^{(2)}(0)$ against $r$ for different values of $\gamma$ and $\bar{n}_m$. For $\bar{n}_m = 0$ the steady state presents super-Poissonian statistics, with large values of $g^{(2)}(0)$, specially for small values of $r$ (actually, $g^{(2)}(0)$ is not defined in the limit $r \to 0$ if $\bar{n}_m = 0$). However, we notice an interesting behavior if $\bar{n}_m$ is increased. For instance, if we take $\bar{n}_m = 10, 100$, we observe much smaller values for $g^{(2)}(0)$, i.e., a strongly sub-Poissonian field state, specially for small values of $r$. This indicates that by increasing the temperature of the reservoir, the system is able to reach a steady state with more pronounced quantum features. This phenomenon can be understood if we take a closer look at the effective master equation (\ref{meae}). For small values of $r$, $\hat{\beta}^2 \approx \hat{b}^2 + 2 r \hat{b}^{\dagger} \hat{b} + r$, so that the first term yields a (strong) two phonon dissipative process and the second and third term preserve the number of phonons, therefore not acting really as a dissipative term. If $\bar{n}_m = 0$, eq. (\ref{meae}) gives us a dynamics in which the mechanical oscillator continuously loses phonons, so that the steady state is the vacuum state, and that is why $g^{(2)}(0)$ diverges for $r \to 0$. However, if $\bar{n}_m \neq 0$ we add a process by which the mechanical oscillator gains phonons from the environment, and the steady state is not the vacuum state anymore.  To explain why it is possible to reach such small values for $g^{(2)}(0)$ we must remember that we have a strong two phonon dissipative process, so that the state of the system converges into a mixture of the states $|n_a = 0,n_b = 0\rangle$ and $|n_a = 0,n_b = 1\rangle$, which may have the strong sub-Poissonian statistics observed. In \cite{nunnenkamp} the authors used a rate equation to explain this phenomenon. This explains the increase in the fidelity observed in figures \ref{fig3c} and \ref{fig3d}. The nonzero population in the one-phonon state increases the fidelity of the steady state with the squeezed one phonon state. For larger values of $r$ a two phonon creation process comes into play and it is not possible to observe sub-Poissonian statistics for the mechanical oscillator anymore. It is important to note that, as sub-Poissonian statistics arises for small $r$, the required modulation amplitude of the mechanical mode, $\epsilon$, is also small. Actually, for $r=0$, where the smallest values for $g_2^{(2)}$ have been found, we have that $\epsilon = 0$.

\begin{figure}
	\centering	
	\subfigure{
		\includegraphics[width=4.2cm]{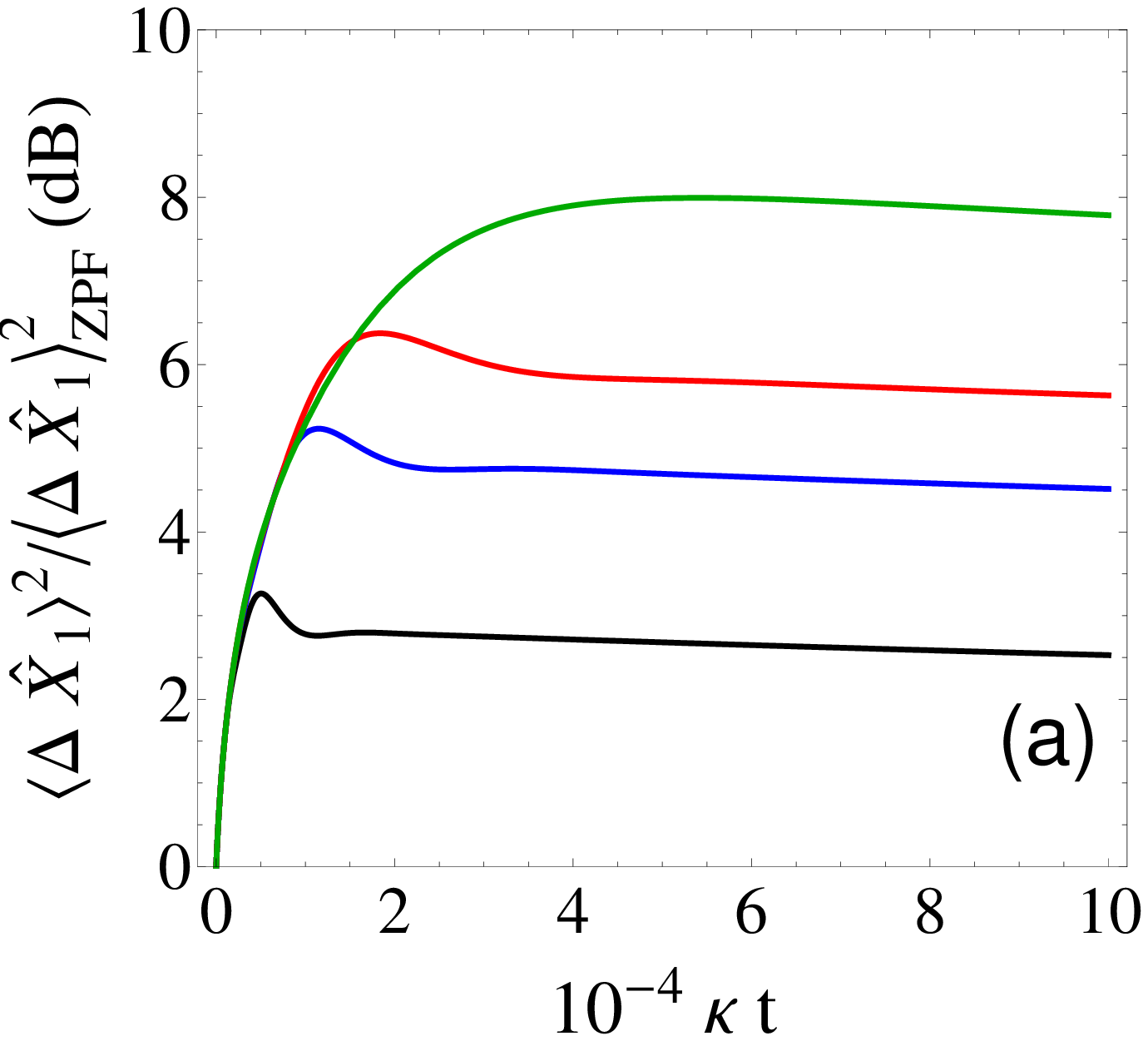}
		\label{fig6a}
	}\hspace{-0.3cm}
	\subfigure{
		\includegraphics[width=4.2cm]{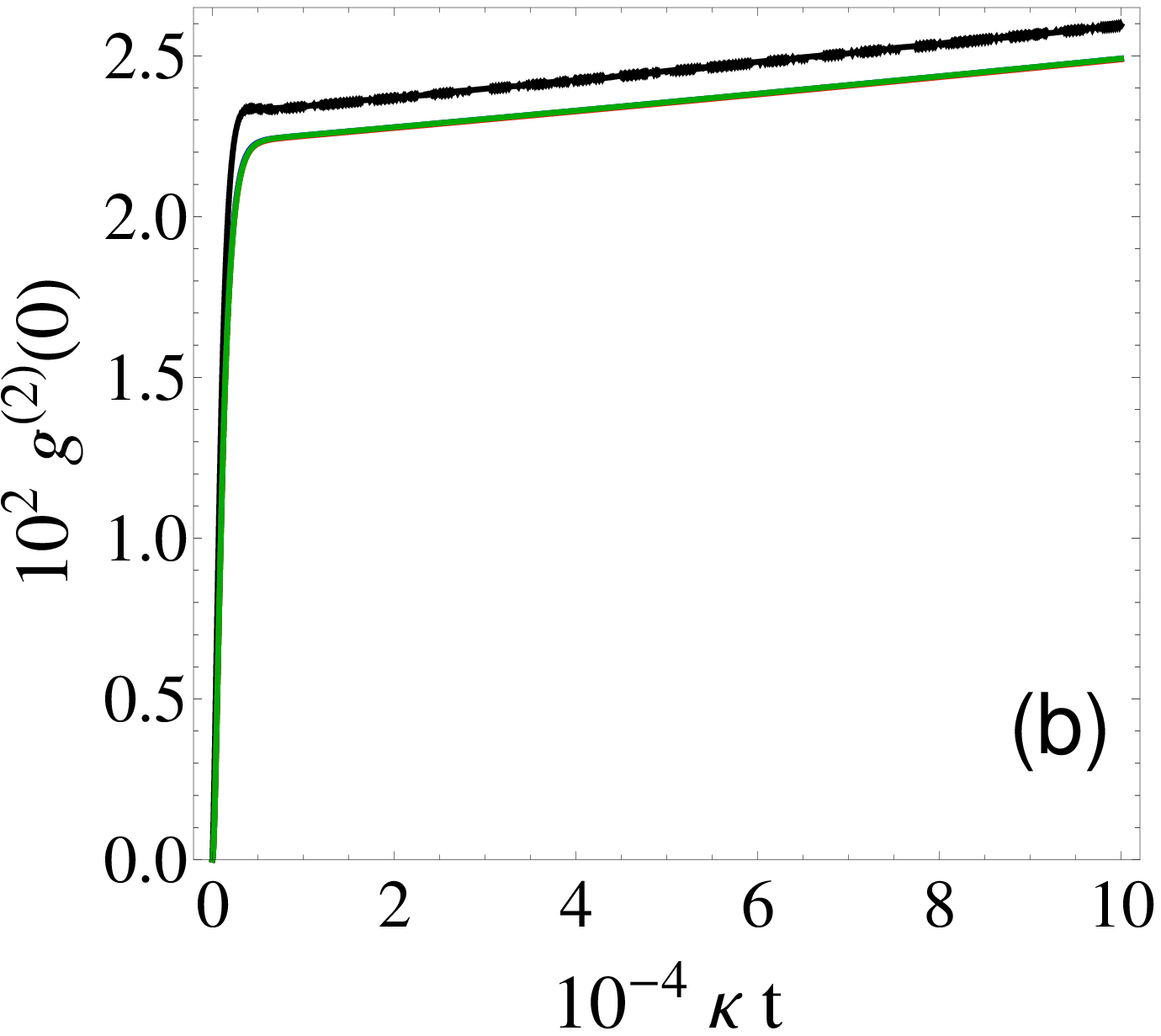}
		\label{fig6b}
	}
	\caption{(a) Quadrature squeezing of $\hat{X}_1$ against $\kappa t$ for $r = 1$ and different values of $\tilde{\omega}_m/\kappa$. The initial state of the system is $|n_a = 0, n_b = 0 \rangle$. (b) $g^{(2)}(0)$ against $\kappa t$ for $r = 0.05$ and different values of $\tilde{\omega}_m/\kappa$. The initial state of the system is $|n_a = 0, n_b = 1 \rangle$. In the black curves $\tilde{\omega}_m = 5 \kappa$, in the blue curves $\tilde{\omega}_m = 20 \kappa$, in the red curves $\tilde{\omega}_m = 50 \kappa$, and in the green curves we take the limit $\tilde{\omega}_m/\kappa \to \infty$. In all curves $\gamma = 10^{-6} \kappa$ and $\bar{n}_m = 0$.}
	\label{fig6}	
\end{figure}

As pointed out in the last section, the results presented here rely on the validity of the rotating wave approximation, which is known to hold if the system operates in the deep resolved sideband regime ($\kappa \ll \tilde{\omega}_m$). However, in realistic situations this condition is hardly fulfilled, and it would be interesting to study the behavior of our system for nonzero values of the ratio $\kappa/ \tilde{\omega}_m$. To this end, we must consider the full (time dependent) Hamiltonian of eq. (\ref{h2}). We have that, if we do not work in the deep resolved sideband regime ($\kappa/\tilde{\omega}_m < 10^{-3}$), the first term of Hamiltonian (\ref{hnr}) may bring a contribution comparable to (or even larger than) the contributions of the resonant terms themselves. To avoid this term and obtain squeezing for moderate values of the sideband parameter, we consider that the mechanical oscillator has also a modulation at the frequency of $4 \tilde{\omega}_m$, which is adjusted to cancel the first term of Hamiltonian (\ref{hnr}). For this situation, we have solved numerically the master equation (\ref{me1}) using the full Hamiltonian, and the time evolution of the squeezing of the $\hat{X}_1$ quadrature and of the $g^{(2)}(0)$, for different value of $\tilde{\omega}_m/\kappa$, is shown in figure \ref{fig6}. In figure \ref{fig6a} the results show that the nonresonant heating of the Bogoliubov mode generated by $\hat{H}_{NR}$ compromises the squeezing of the $\hat{X}_1$ significantly. However, it is still possible to observe squeezing above the $3$ $db$ limit even for  $\tilde{\omega}_m/\kappa \approx 5$. On the other hand, we have found that the influence of the nonresonant terms on the generation of states with sub-Poissonian statistics is considerably smaller. As one can note in figure \ref{fig6b}, the blue curve ($\tilde{\omega}_m/\kappa = 20$) and red curve ($\tilde{\omega}_m/\kappa = 50$) can not be distinguished from the green curve ($\tilde{\omega}_m/\kappa \to \infty$). We believe that this small effect is due to the fact that sub-Poissonian statistics appears only for small values of $r$, and most of the nonresonant terms are proportional to $\sinh{r}$ and $\sinh^2{r}$, which are also small. We also believe that effects of the nonresonant terms are minimal because in the quadratic case they oscillate at frequencies $2 \tilde{\omega}_m$ or $4 \tilde{\omega}_m$, differently from what usually occurs in a linear optomechanical system, for which the frequency of nonresonant terms is $\tilde{\omega}_m$.

Regarding the experimental feasibility of our proposal, we consider here the realistic parameters of a quadratic optomechanical system reported in \cite{sankey}: $\omega_m \approx 10^{6} Hz$, $\kappa \approx 10^{5} Hz$, $\gamma \approx 0.1 Hz$, $g \approx 10^{-4} Hz$, and the mass of the mechanical oscillator is $m = 30 ng$. For a laser with power $P_{in} = 5 \mu W$, the effective optomechanical coupling is $g \alpha \approx 0.76 Hz \approx 10^{-5} \kappa$ \cite{tan}. However, the effective coupling $g \alpha$ is far below the value used here, $10^{-2} \kappa$. A possible way to get around this would be to increase $P_{in}$; nevertheless, this would result in a very high value for $\epsilon$ and probably would lead the optomechanical system to instability (actually, if one is interested in the observation of sub-Poissonian statistics for $r = 0$, it would be possible to increase $P_{in}$, as no modulation of the mechanical oscillator is needed in this case). A solution to that problem would be to increase the value of $g$ by five orders of magnitude. Indeed, in \cite{flowers-jacobs} the authors have measured $g \approx 5 Hz$ but, unfortunately, the optomechanical system does not work in the resolved sideband regime. Considering an optomechanical system with $g \approx 10 Hz$, then the amplitude of the parametric driving must be $\epsilon = 8 \kappa/\omega_m$, and therefore, a system working in the deep resolved sideband regime would require a small driving of the mechanical oscillator. Another question that must be considered is the tolerance of our purpose regarding the condition imposed on the driving amplitudes, $\epsilon = - 2 \lambda$. Our simulations show that the relation $\epsilon = - 2 \lambda$ must be satisfied with a very high precision, about $10^{-4}\%$, in order to observe the desired effects. %In short, we believe that future realizations of optomechanical systems with larger quadratic optomechanical couplings, and working in the resolved sideband regime, would allow the experimental realization of our proposal. 

\section*{VI. CONCLUSIONS}
In this work we have studied an optomechanical system with a purely quadratic optomechanical coupling, whose mechanical mode is being parametrically driven and whose optical mode is being pumped by three coherent fields. We have shown that if the frequencies and amplitudes of both parametric and optical drivings are chosen properly, the optomechanical system acts as a mechanical oscillator coupled to an engineered reservoir. Under these circumstances, and in the absence of mechanical damping, the squeezed vacuum state and the squeezed one phonon state (or any mixture of them) are dark states of the dynamics. This was  confirmed by our numerical results. We made a detailed analysis of time evolution of the system for $\gamma \neq 0$, and the results have shown two time scales, one determined by the decay rate of the optical mode $\kappa$ and the coupling constant $\mathcal{G}$, and other determined by the decay rate of the mechanical mode $\gamma$, and $\bar{n}_m$. Our results have shown that if $t \gg \mathcal{G}^2/\kappa$ (if the adiabatic elimination is valid) and $t \ll [(\bar{n}_m + 1) \gamma]^{-1}$, the state of the system is very well approximated by the states $|0 \rangle_a \otimes |\xi, 0 \rangle_b$ or $|0 \rangle_a \otimes |\xi, 1 \rangle_b$, depending on the initial state of the system. We have shown that even with $\gamma \neq 0$ the system still presents steady states with large degrees of squeezing (above $7.9$ $dB$), and strong sub-Poissonian statistics ($g^{(2)}(0) < 2 \times 10^{-2}$). We have also analyzed the effect of a nonzero sideband parameter $\kappa/\tilde{\omega}_m$ on the generation of squeezed states, and found that in general it has deleterious effects on squeezing, although it is still possible to observe reduction of noise above $3$ $dB$ in the moderate resolved sideband regime ($\tilde{\omega}_m/\kappa \approx 5$). Nevertheless, the effects of the nonresonant terms on the generation of sub-Poissonian states were small even for $\tilde{\omega}_m/\kappa \approx 5$ and can hardly be noted for larger values of the ratio $\tilde{\omega}_m/\kappa$.

\begin{acknowledgments}
This work was supported by the São Paulo Research Foundation (FAPESP)(project No. 2012/10476-0), by the National Council for Scientific and Technological Development (CNPq), by the Optics and Photonics Research Center (CePOF) and the Brazilian National Institute for Science and Technology of Quantum Information (INCT-IQ).
\end{acknowledgments}

\section*{Appendix}
\begin{figure}[H]
	\centering	
	\includegraphics[height=3.825cm]{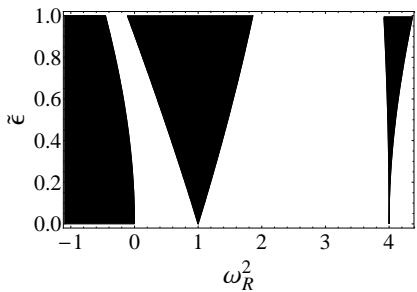}
	\caption{Stability diagram of the Mathieu equation. In the white areas there are just stable solutions while in the black areas there is at least one unstable solution.}
	\label{fig7}	
\end{figure}

In this Appendix we discuss the possibility of unstable solutions of eq. (\ref{sigma0}). Let us consider first the parametric oscillator equation of motion:
\begin{equation}
	\ddot{x} + \gamma \dot{x} + \bigl[\omega^2 + \epsilon \cos(\Omega t) \bigr] x = 0,
	\label{em1}
\end{equation}
where $\epsilon$ and $\Omega$ are the amplitude and frequency of the modulation, respectively, and $\gamma$ is the decay rate of the oscillator. Defining the dimensionless parameters,
\begin{equation}
	\begin{array}{ll}
		\tilde{t} = \frac{\Omega t}{2}, \\
		\tilde{\omega} = \frac{2 \omega}{\Omega},
	\end{array} 
	\quad
	\begin{array}{ll}
		\tilde{\epsilon} = \frac{2 \epsilon}{\Omega^2}, \\
		\tilde{\gamma} = \frac{2 \gamma}{\Omega},
	\end{array} 	
	\label{em1.1}
\end{equation}
we can write the equation of motion in the following way,
\begin{equation}
	\ddot{x} + \tilde{\gamma} \dot{x} + \bigl[\tilde{\omega}^2 + 2 \tilde{\epsilon} \cos(2 \tilde{t}) \bigr] x = 0.
	\label{em2}
\end{equation}
Defining $x = y \exp(-\tilde{\gamma} \tilde{t}/2)$ and substituing in eq. (\ref{em2}), we obtain, 
\begin{equation}
	\ddot{y} + \left[ \omega_R^2 + 2 \tilde{\epsilon} \cos(2 \tilde{t}) \right] y = 0.
	\label{em3}
\end{equation}
This is a Mathieu equation with a renormalized angular frequency $\omega^2_R = \tilde{\omega}^2 - \tilde{\gamma}^2/4$. Using the results of the Floquet theory for differential equations, it is possible to show that the Mathieu equation has unstable solutions \cite{wang}. In figure \ref{fig7} we have the stabity diagram of the Mathieu equation, where in the dark areas at least one of the solutions is unstable. If $\Omega = 2 \omega$, then $\omega_R = 1 - (\gamma/\omega)^2/4$ and $\tilde{\epsilon} = \epsilon/2$, and for an oscillator with a high quality factor, instability may arise even for small $\tilde{\epsilon}$. However, if $\Omega = 4 \omega$, then $\omega_R = 1/2 - (\gamma/\omega)^2/16$ and $\tilde{\epsilon} = \epsilon/8$, and for small enough $\epsilon$, the stability of the system is guaranteed. Given that the solution of equation \ref{em1} is $x = y \exp(-\tilde{\gamma} \tilde{t}/2)$, then $\lim_{t \to \infty} x(t) = 0$.

\end{document}